\documentclass[aps,prb,twocolumn,superscriptaddress,longbibliography]{revtex4-1}
\usepackage[colorlinks=true,citecolor=blue,linkcolor=blue,breaklinks=true]{hyperref}

\usepackage{color,graphicx}
\usepackage{bm}
\usepackage{amsmath}
\usepackage{amssymb}
\usepackage{times}

\begin{document}
	
\newcommand {\beq} {\begin{equation}}
\newcommand {\eeq} {\end{equation}}
\newcommand {\bqa} {\begin{eqnarray}}
\newcommand {\eqa} {\end{eqnarray}}
\newcommand {\ca} {\ensuremath{c^\dagger}}
\newcommand {\ba} {\ensuremath{b^\dagger}}
\newcommand {\Ma} {\ensuremath{M^\dagger}}
\newcommand {\psia} {\ensuremath{\psi^\dagger}}
\newcommand {\fbar} {\ensuremath{\bar{f}}}
\newcommand {\psita} {\ensuremath{\tilde{\psi}^\dagger}}
\newcommand{\lp} {\ensuremath{{\lambda '}}}
\newcommand{\A} {\ensuremath{{\bf A}}}
\newcommand{\Q} {\ensuremath{{\bf Q}}}
\newcommand{\kk} {\ensuremath{{\bf k}}}
\newcommand{\qq} {\ensuremath{{\bf q}}}
\newcommand{\kp} {\ensuremath{{\bf k'}}}
\newcommand{\rr} {\ensuremath{{\bf r}}}
\newcommand{\rp} {\ensuremath{{\bf r'}}}
\newcommand {\ep} {\ensuremath{\epsilon}}
\newcommand{\nbr} {\ensuremath{\langle r r' \rangle}}
\newcommand {\no} {\nonumber}
\newcommand{\up} {\ensuremath{\uparrow}}
\newcommand{\dn} {\ensuremath{\downarrow}}
\newcommand{\rcol} {\textcolor{red}}
\newcommand{\bcol} {\textcolor{blue}}
\newcommand{\lt} {\left}
\newcommand{\rt} {\right}

\title{Thermal effects on collective modes in disordered $s$-wave superconductors}
\author{Abhisek Samanta}\email{abhiseks@campus.technion.ac.il}
\affiliation{ Physics Department, Technion, Haifa 32000, Israel}
\author{Anirban Das}
\affiliation{ School of Physical Sciences, Indian Association for the Cultivation of Science, Jadavpur, Kolkata 700032, India}
\author{Nandini Trivedi}
\affiliation{ Department of Physics, The Ohio State University, Columbus, Ohio, USA 43201}
\author{Rajdeep Sensarma}\email{sensarma@theory.tifr.res.in}
\affiliation{Department of Theoretical Physics, Tata Institute of Fundamental Research, Mumbai 400005, India.}

\date{\today }

\begin{abstract}
We investigate the effect of thermal fluctuations on the two-particle spectral function for a disordered $s$-wave superconductor in two dimensions, focusing on the evolution of the collective amplitude and phase modes. We find three main effects of thermal fluctuations: (a) the phase mode is softened with increasing temperature reflecting the decrease of superfluid stiffness; (b) remarkably, the non-dispersive collective amplitude modes at finite energy near ${\bf q}=[0,0]$ and ${\bf q}=[\pi,\pi]$ survive even in presence of thermal fluctuations in the disordered superconductor; and (c) the scattering of the thermally excited fermionic quasiparticles leads to low energy incoherent spectral weight that forms a strongly momentum-dependent background halo around the phase and amplitude collective modes and broadens them. Due to momentum and energy conservation constraints, this halo has a boundary which disperses linearly at low momenta and shows a strong dip near the $[\pi,\pi]$ point in the Brillouin zone. 
\end{abstract}
\pacs{XXXX}

\maketitle

\section{Introduction}
The quantum phase transition between superconducting and insulating phases of two dimensional films, driven by increasing disorder~\cite{Goldman, Sacepe, Gantmakher, Avishai, Shahar, Nandini3}, provides a paradigm for the complex interplay of interaction and localization~\cite{Feigelman, Mirlin, Nandini1, Nandini2, Gershenson}. The single particle fermionic spectrum remains gapped throughout the transition~\cite{Pratap, Nandini1} and the fluctuations of the local superconducting order parameter, that describe the phase (Goldstone) and amplitude (Anderson-Higgs) collective modes, are the key low energy excitations that drive this phase transition~\cite{BlochHiggs, HiggsExpt, Sherman}.

The Higgs mode in superconductors has been an active area of research~\cite{Anderson-Higgs, HiggsRevShimano, HiggsRevPekker} for a long time. While the observation of the Higgs particle in particle colliders~\cite{ATLAS, CMS} has been hailed as one of the recent successes in that field, the corresponding mode has not been observed in a clean superconductor. This is due to the fact that in a clean superconductor, this mode sits at the two-particle continuum threshold and is damped~\cite{HiggsRevPekker}. Early predictions~\cite{VermaLittlewood} that the mode can be seen as a subgap feature in systems with accompanying charge density order has recently been experimentally verified~\cite{HiggsRaman}. The area has also received a lot of attention due to clean observation of the Higgs mode in a charge neutral ultracold atomic system~\cite{BlochHiggs} near the superfluid-insulator transition. More recently, observation of low energy optical spectral weight in disordered superconductors~\cite{Sherman} close to a disorder driven superfluid-insulator transition has led to a conjecture that this weight is due to the Higgs mode, based on earlier work on optical conductivity in clean systems~\cite{AuerbachPodolsky1, AuerbachPodolsky2}. More recent theoretical work~\cite{Benfatto1, Benfatto2} has looked at the question of the contribution of collective modes to optical conductivity in disordered superconductors at zero temperature.

In an earlier work~\cite{HiggsAbhisek}, we had studied the evolution of two-particle pair spectral functions in a disordered superconductor and presented their full momentum and frequency dependence as a function of disorder at zero temperature. We had found the expected (a) continuum of two-particle excitations, above an energy threshold equal to twice the single-particle gap, and (b) linearly dispersing low energy collective modes. In addition, surprisingly, we found additional spectral weight at finite energies below the two-particle continuum in the long wavelength limit. The weight in this non-dispersive feature, which was spectrally separated from the linearly dispersing collective modes, increased with increasing disorder strength. We were able to correlate this non-dispersive spectral weight with the Higgs mode and the low energy Higgs weight was concentrated in this additional spectral feature in the two-particle pair spectral function. One obvious question is how does that picture change in the presence of temperature? It is crucial to understand the combined role of both thermal and quantum fluctuations in order to make connections with experimental data.  

We had also found that at arbitrarily weak disorder, the zero momentum Anderson-Higgs mode that sits at the threshold of the two-particle continuum, shifts {\it non-perturbatively} within the two-particle gap. This subgap feature of the Anderson-Higgs mode is distinguishable from the low energy phase divergence at all disorder values, making it a possible candidate to observe in energy resolved spectroscopies. Therefore, the natural question that arises is: what is the fate of the Anderson-Higgs mode as a function of temperature? Is it still possible to separate this mode from the phase fluctuations at finite temperatures?
We address these important questions in this work. 


Our theoretical approach involving functional integrals allows us to investigate the different contributions of the amplitude, the phase, and amplitude-phase mixing fluctuations to the two-particle spectral function at finite temperature. The key features of our analysis are the following: i) We obtain the evolution of the Anderson-Higgs and the Goldstone mode as a function of both temperature and disorder. ii) We find that small thermal fluctuations induce additional low energy incoherent spectral weight that forms lobes below the two-particle gap, which compete with the collective mode structure in the amplitude sector, but keep the phase sector mostly unaffected. iii) In presence of weak disorder, the subgap Anderson-Higgs mode can be observed separately from the low energy phase pile-up in an energy resolved way even in presence of moderately high temperatures, thereby making it a robust feature of disordered superconductors. We note that an alternative approach based on an effective classical Monte Carlo has been used to treat disordered superconductors at finite temperature~\cite{Tarat}, but it does not provide momentum resolved information about the amplitude and phase fluctuations.

\begin{figure}
	\includegraphics[width=0.95\columnwidth]{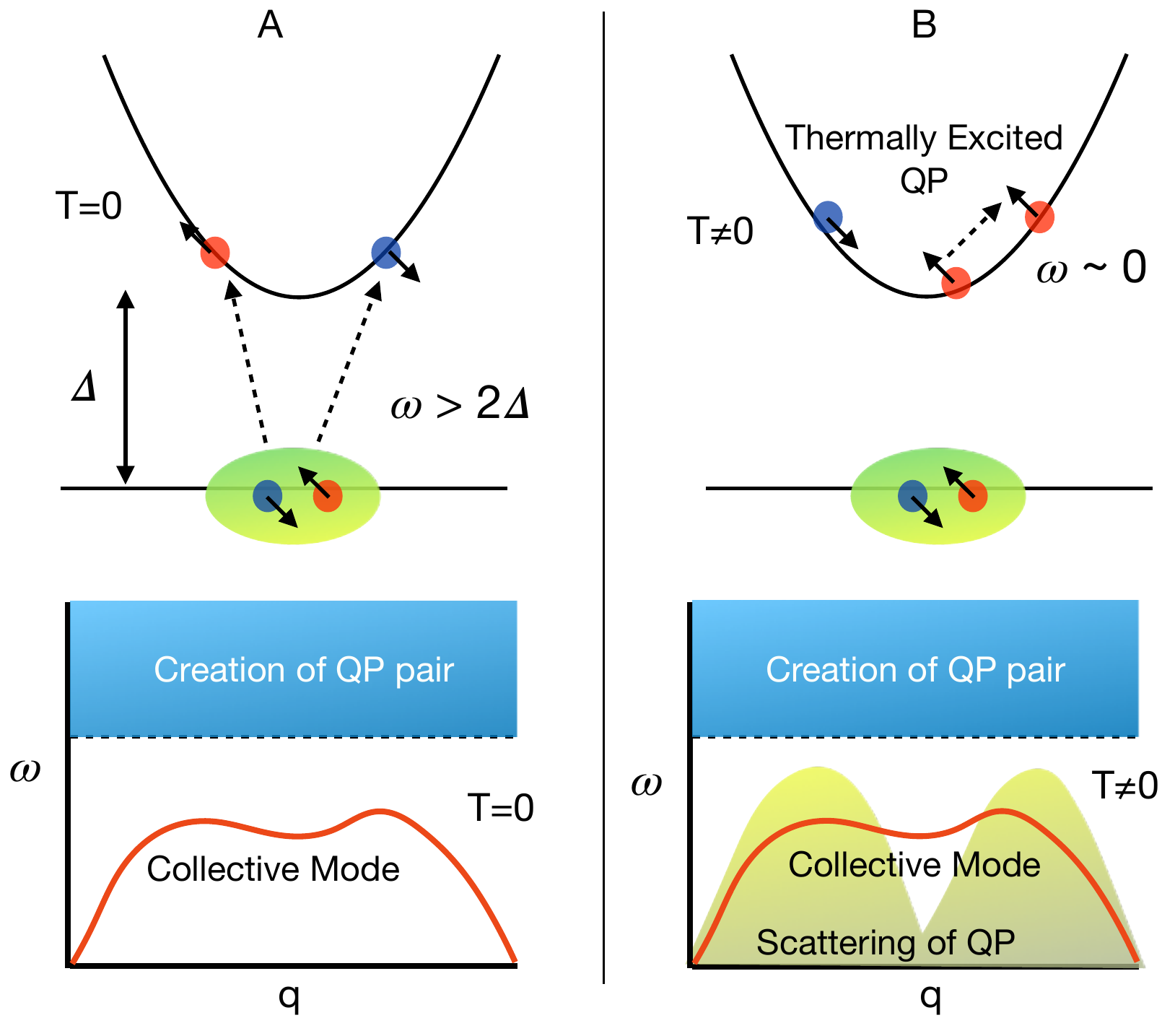}
	\caption{ (A) At $T=0$, the breaking of the Cooper pair into two quasiparticles leads to a two-particle continuum in the pair spectral weight  with a threshold of $2\Delta$. The lower part of the figure depicts a schematic of the features of the pair spectral function at $T=0$. (B) At finite temperature, the scattering of thermally excited quasiparticles leads to additional low energy incoherent spectral weight. This leads to a halo behind the collective modes, as shown in the lower part of the figure. }
	\label{Fig:cartoon}
\end{figure}

The two-particle continuum at $T=0$ is formed microscopically by breaking up a Cooper pair into a pair of single-particle excitations, as shown in Fig.~\ref{Fig:cartoon} (A) and requires a threshold energy of twice the single-particle gap. On the other hand, the collective mode is better described in terms of the long wavelength fluctuations of the amplitude and phase of the condensate of the Cooper-pairs (i.e. the order parameter). As temperature is raised, changes in the pair spectral function occur by two processes: (a) The collective mode dispersion flattens as the superfluid stiffness is reduced at finite temperatures due to thermally excited quasiparticles. (b) Additionally, another incoherent continuum is formed due to scattering of these thermally populated quasiparticles, as shown in Fig.~\ref{Fig:cartoon} (B). This leads to a low energy diffuse background weight and consequent broadening of the collective modes. In a clean system, the incoherent thermal excitations occur only below an upper energy cut-off $\epsilon_{cut-off}(q)$ determined by energy and momentum conservation in the scattering process.
$\epsilon_{cut-off}(q)$ varies linearly at long wavelengths and shows a prominent dip around the commensurate wave-vector $[\pi,\pi]$.

In a weakly disordered system, the behavior of the energy-cutoff and its momentum-dependence continues to hold with small corrections. As a result, the non-dispersive spectral weight observed at finite subgap energy at long wavelengths remains sharp at finite temperatures for weakly disordered systems. For strongly disordered systems, the constraint due to momentum conservation in a scattering process is no longer applicable, and we find the incoherent spectral weight as a diffused halo without sharply defined boundaries.
Since the low energy weight in the diffuse halo comes from the scattering of thermally excited quasiparticles, it is exponentially suppressed at low temperatures, and significant weight develops only when a fraction of the critical temperature $T_c$ is approached.

In disordered systems, the presence of a reasonably sharp threshold of the incoherent weight at long wavelengths leads to a clear visibility of the long wavelength finite energy weight in the Higgs spectrum. This spectral feature, which was seen clearly in the $T=0$ calculations~\cite{HiggsAbhisek}, thus survives thermal fluctuations in the system. This is a key insight that we obtain from these calculations. 

The rest of the paper is organized as follows: In Section II we discuss the model Hamiltonian for disordered superconductors and the finite temperature mean-field theory. 
In Section III, we first discuss the technical details of the finite temperature gaussian fluctuation calculation, before turning our attention to the pair spectral function in clean systems at finite temperature in Section III A. In Section III B, we focus on the finite temperature evolution of the pair spectral function in  the disordered system. Finally, we conclude with a brief overview of our calculation and key results.
\begin{figure*}
	\includegraphics[width=0.9\textwidth]{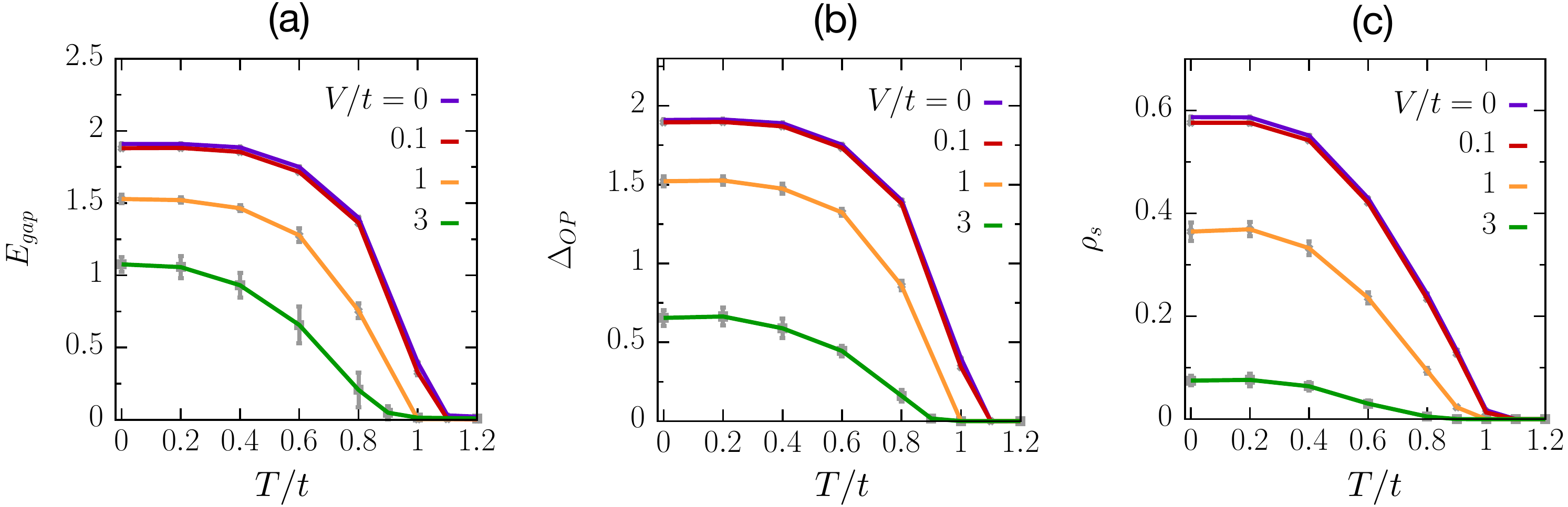}
	\caption{ (a) Single-particle excitation gap  $E_{gap}$, (b) Superconducting order parameter $\Delta_{OP}$,  and (c) superfluid stiffness $\rho_s$ as a function of temperature $T$, obtained from finite $T$ BdG calculation. We present data for clean system ($V=0$) and three disorder values, $V=0.1~t$, $t$ and $3~t$. $E_{gap}$ and $\Delta_{OP}$ vanish at $T_c=1.1~t$ in the clean system. All three parameters, $E_{gap}, \Delta_{OP}$ and $\rho_s$ have reduced significantly at $T=0$, while the decrease in mean-field $T_c$ is much smaller. The data have been obtained on a $24\times24$ square lattice and averaged over 15 disorder realizations.}
	\label{Fig:mft}
\end{figure*}

\section{Mean-field theory at finite temperatures~\label{model}}
We study the attractive Hubbard model on a square lattice in the presence of on-site non-magnetic impurities. The Hamiltonian for the model is given by,
\begin{equation}
\label{eq:Ham}
H=-t\sum_{\nbr\sigma}(\ca_{r\sigma}c_{r' \sigma} + h.c )-U\sum_rn_{r\up}n_{r\dn}+\sum_r (v_r-\mu)n_r
\end{equation}
where $c^{\dagger}_{r\sigma}(c_{r\sigma})$ is the creation (annihilation) operator for an electron with spin $\sigma$ on site $r$, and $\mu$ is the chemical potential. Nearest neighbour hopping between two electrons is governed by $t$, and $U$ is the attractive interaction between two electrons on the same site which leads to Cooper pairing. Here, $v_r$ is an on-site random potential, which is drawn independently on every site $r$ from a uniform distribution of zero mean and width $V$, i.e. $v_r\in [-V/2,V/2]$. Therefore, $V$ corresponds to the strength of the disorder.
This model has been studied previously~\cite{Nandini1} at zero temperature within a spatially inhomogeneous Bogoliubov de-Gennes (BdG) mean-field theory. More recently the two-particle spectral functions in this model at $T=0$ have been studied within a gaussian expansion around the BdG solution~\cite{HiggsAbhisek}. In this section we investigate the mean-field theory at finite temperatures, while later sections will be devoted to considering the fluctuations around the mean-field theory at finite temperatures.


Within a functional integral formalism, the partition function for the model is given by,
\begin{equation}
Z = \int D[\bar{f}_\sigma,f_\sigma]e^{-S[\bar{f}_\sigma,f_\sigma]}.
\end{equation}
Here the imaginary time ($\tau$) action $S$ in terms of the fermion fields ($\bar f_{\sigma}(r,\tau), f_{\sigma}(r,\tau)$) is given by  
\begin{eqnarray}
\label{eq:saddlept}
S &=& \int_0^\beta d\tau \sum_{rr',\sigma} \bar{f}_{\sigma}(r,\tau) 
\left[ \partial_\tau\delta_{rr'}+ H^0_{rr'}\right]f_{\sigma}(r',\tau) \no \\ &&
-U\sum_r\bar{f}_{\up}(r,\tau)\bar{f}_{\dn}(r,\tau)f_{\dn}(r,\tau)f_{\up}(r,\tau), 
\end{eqnarray}
where $\beta=1/T$ and the single-particle Hamiltonian $H^0_{rr'}=-t\delta_{\langle rr'\rangle}-(\mu-v_r)\delta_{rr'}$. We introduce two Hubbard-Stratonovich auxiliary fields, $\Delta(r,\tau)$ that couples to the particle-particle channel ($\bar{f}_\up(r,\tau)\bar{f}_\dn(r,\tau)$), and the field $\xi(r,\tau)$ that couples to the density channel ($\bar{f}(r,\tau)f(r,\tau)$), in order to construct a quadratic theory in the fermion fields. Integrating out the fermions, and considering a static but spatially varying saddle point profile of the auxiliary fields, $\Delta(r,\tau) = \Delta_0(r)$ and $\xi(r,\tau) = \xi_0(r)$, lead to the BdG mean-field theory. The BdG self-consistency
equations at finite $T$ are given by,
\begin{gather}
\label{eq:selfc1}
\Delta_0(r) = U\sum_n u_n(r)v_n^*(r)\lt(1-2F_{n}(T)\rt), \\
\xi_0(r) = U\sum_n (1-2F_n(T))|v_n(r)|^2 + F_{n}(T)|u_n(r)|^2, \\
\text{and}~~ \langle n\rangle = \frac{2}{N_s}\sum_r \frac{\xi_0(r)}{U} ,
\label{eq:selfc2}
\end{gather}
where $\langle n\rangle$ is the average density of electrons in the system with $N_s$ number of sites. Here $[u_n(r),v_n(r)]$ are the eigenvector of the BdG matrix corresponding to the eigenvalue $E_n$ and $n$ runs over positive eigenvalues ($E_n>0$) only. The Fermi function at a temperature $T$ is given by $F_{n}(T) = \frac{1}{e^{ E_{n}/T}+1}$. We solve the BdG self-consistency equations (Eqn.~\ref{eq:selfc1}-\ref{eq:selfc2}) on a $24\times 24$ square
lattice with an interaction strength $U/t=5$ and at an average fermion density $\langle n \rangle=0.875$. We consider 15 disorder realizations for each disorder.

Before we discuss the results of the mean-field theory at finite temperatures, we note the main features
of the mean-field theory at zero temperature for disordered
superconductors~\cite{Nandini1, HiggsAbhisek}: (i)  The distribution of the
local pairing amplitude
evolves from a sharp distribution around an average value for low disorder to a broad
distribution with peaks around zero for large disorder, which indicates the destruction of superconductivity. (ii) The
distribution of local densities evolves from a sharp unimodal
distribution at low disorder to a broad bimodal distribution at large
disorder. This indicates the formation of superconducting puddles or
patches in the background of non-superconducting regions at large
disorder.  (iii) The formation of superconducting patches is further
confirmed by the spatial distribution of the
pairing amplitude that shows cluster formation on the scale of the coherence length in typical disorder configurations. (iv) The
single-particle gap remains finite and large at strong disorder, while
the average order parameter and the superfluid density both decrease
monotonically at large disorder. We will next compare and contrast these features to the behavior at finite temperatures.
\begin{figure*}
	\includegraphics[width=0.9\textwidth]{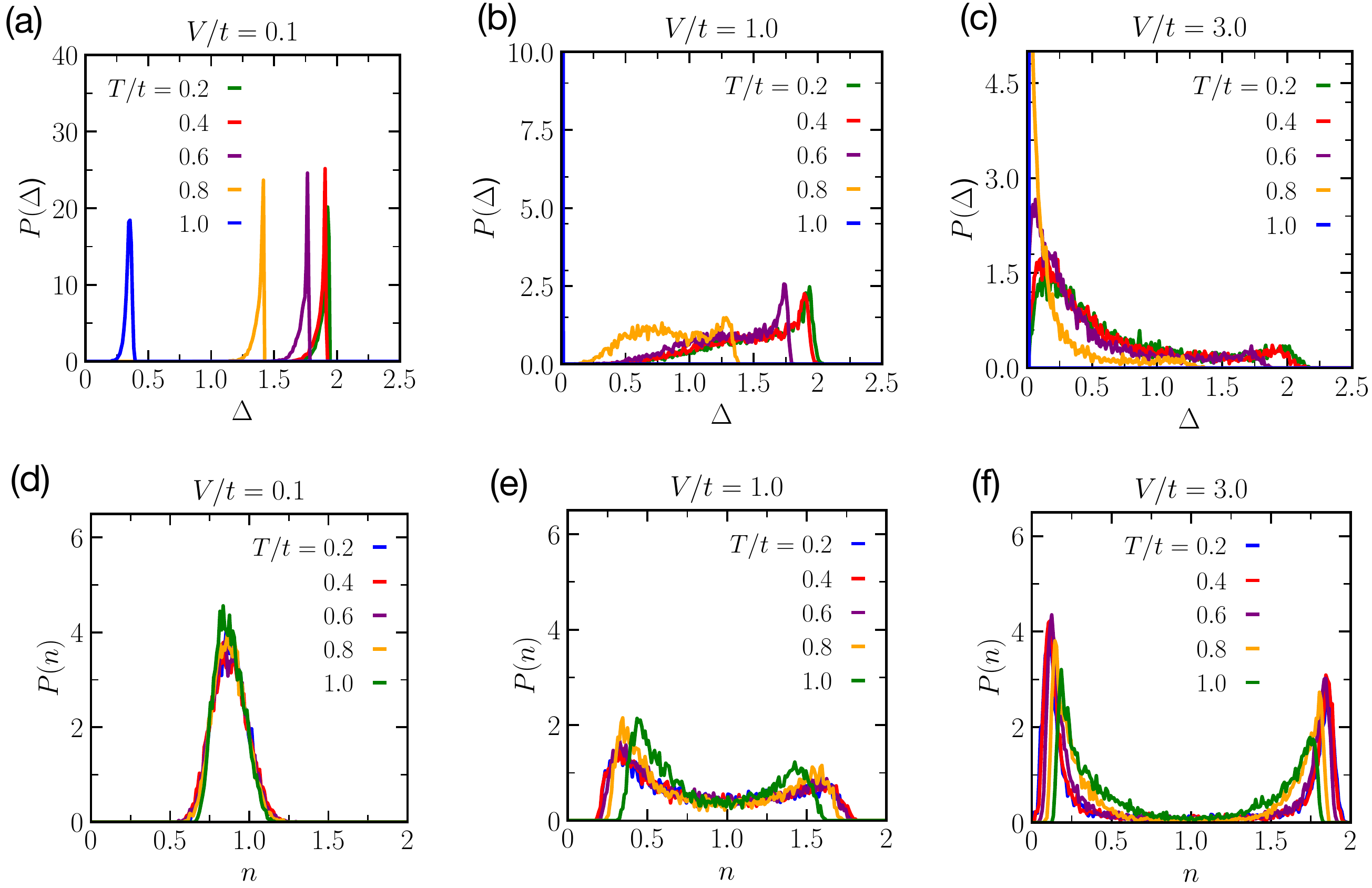}
	\caption{ Distribution of (a)-(c) local superconducting pairing amplitude $\Delta(r)$ and (d)-(f) local density $n(r)$ obtained from finite $T$ BdG calculation for different values of disorder $V$ and temperature $T$. With increasing disorder, $P(\Delta)$ changes from a sharp distribution around the average $\Delta$ to a broad distribution with peaks around 0. For a fixed $V$, the shape of $P(\Delta)$ does not change much with temperature, while the distribution shifts towards the lower value of $\Delta$. On the other hand, $P(n)$ becomes bimodal around the average density $\langle n\rangle=0.875$ in presence of strong disorder, and with increasing the temperature, the distribution becomes narrower.}
	\label{dist}
\end{figure*}

\medskip
\noindent {\it{Temperature dependence of single-particle gap}}:
Fig.~\ref{Fig:mft} (a) shows the single-particle gap in
the system as a function of temperature. The gap
for a clean superconductor ($V=0$) vanishes around
$T_c=1.1~t$. 
Note
that while the $T=0$ gap has reduced by a factor of $2$ between the
clean case and $V=3~t$, the decrease in the mean-field $T_c$ is much
smaller. A similar trend is seen in Fig.~\ref{Fig:mft} (b) where we plot the average pairing amplitude $\Delta_{OP}$ (averaged over sites and over disorder realizations) as a function of temperature for different values of disorder. 
Once again we note that while $\Delta_{OP}(T=0)$ reduces by a factor of four as we go from $V=0$ to $V=3~t$, $T_c$ only changes from $1.1~t$ to $0.9~t$. These two trends taken together show that within the mean-field theory disorder is much more effective at reducing/killing superconductivity at $T=0$ compared to its effect on reducing $T_c$ of the system. 

\medskip
\noindent {\it{Temperature dependence of superfluid stiffness}}: A similar trend is seen in the temperature variation of the superfluid stiffness $\rho_s$ (see Appendix~\ref{app:SFstiff}), which is plotted in Fig.~\ref{Fig:mft} (c) with increasing disorder. 
While the $T=0$ value of $\rho_s$ reduces by a factor of $6$ as the disorder is ramped up from the clean case to $V=3~t$, the transition temperature $T_c$ only changes from $1.1~t $ to $0.9~t$. We note that in two dimensions, the finite temperature transition will be a BKT type transition, with a transition temperature lower than the mean-field $T_c$.

\medskip
\noindent {\it{Distributions}}:
It is useful to look at how the distribution of the local order parameter $\Delta(r)$ and the local density $n(r)$ changes with temperature and disorder strength. In Fig.~\ref{dist} (a) and (b), we plot the distribution of $\Delta(r)$ for $V=0.1~t$ and $V=t$ respectively. Each plot shows the distribution for a range of temperatures. 
In each of these cases, we see that the shape of the distribution does not change much with temperature, although the distribution shifts to lower values of $\Delta$, consistent with the decrease of $\Delta$ with temperature. In Fig.~\ref{dist} (c), we see a similar trend with a pile-up around $\Delta=0$. Note that for $V=t$ and $V=3~t$, $T=t$ is above $T_c$ and we simply get all the weight at $\Delta=0$. We plot the distribution of local densities for $V=0.1~t$, $V=t$ and $V=3~t$ in Fig.~\ref{dist}
(d), (e) and (f) respectively. The density distribution goes from a unimodal distribution at low disorder to a bimodal distribution at high disorder. At all values of disorder, the distribution narrows with increasing temperature, with the effect clearly visible at large disorder strengths. At large disorder, the bimodal distribution comes from the formation of superconducting and non-superconducting patches. Increasing temperature leads to smoother density profile between the patches and hence to a narrowing of the density distributions.

\section{Gaussian fluctuations and pair spectral functions}
The primary motivation of this work is to understand how the fluctuations around the mean-field theory that dominate the two-particle pair spectral function at low energies, evolve with temperature in a disordered superconductor. To this end, we include the spatio-temporal fluctuations of the $\Delta$ field through 
\begin{equation}
\Delta(r,\tau)=(\Delta_0(r) + \eta(r,\tau))e^{i\theta(r,\tau)},
\end{equation}
where $\eta(r,\tau)$ and $\theta(r,\tau)$ 
are the amplitude and the phase fluctuations respectively around the BdG saddle point
solution $\Delta_0(r)$. We expand the action to second order in the fluctuations to obtain the gaussian action $S_G$ corresponding to the fluctuations of the order parameter at finite temperature $T$ (for $T=0$, see Ref.~\onlinecite{HiggsAbhisek}) 
\begin{widetext}  
	\begin{equation}
	S_{G} = \sum_{rr'}\sum_{\omega_m} \left(\begin{array}{cc}\eta(r,\omega_m)&
	\theta(r,\omega_m)
	\end{array}\right)
	\left(\begin{array}{cc}
	{D^{-1}}_{11}(r,r',\omega_m) &
	{D^{-1}}_{12}(r,r',\omega_m) \\
	{D^{-1}}_{21}(r,r',\omega_m) &
	{D^{-1}}_{22}(r,r',\omega_m) 
	\end{array}\right)
	\left(\begin{array}{cc}
	\eta(r',-\omega_m) \\
	\theta(r',-\omega_m)
	\end{array}\right),
	\end{equation}
\end{widetext}
where $\omega_m=(2m)\pi/\beta$ is the bosonic Matsubara
frequency. We analytically continue from Matsubara to real frequencies to construct the real time inverse fluctuation propagators. We note that our formalism does not suffer from issues of numerical analytic continuation. 
\begin{widetext}
	The inverse fluctuation propagator corresponding to the amplitude fluctuation, $D^{-1}_{11}$ is given by,
	\begin{eqnarray}
	{D^{-1}}_{11}(r,r',\omega) 
	&=& \frac{1}{U}\delta_{rr'} + \frac{1}{2}\sum_{E_{n},E_{n'}>0}
	f^{(1)}_{nn'}(r)f^{(1)}_{nn'}(r') \chi_{nn'}(\omega) + 
	\frac{1}{2}\sum_{E_{n},E_{n'}>0}
	{f^{(2)}_{nn'}}(r)f^{(2)}_{nn'}(r') \zeta_{nn'}(\omega),
	\end{eqnarray}
	where 
	\begin{eqnarray}
	f^1_{nn'}(r) &=& \left[u_n(r)u_{n'}(r)-v_n(r)v_{n'}(r)\right],~~ \text{and}  
	~~ 
	f^2_{nn'}(r) = \left[u_n(r)v_{n'}(r)+v_n(r)u_{n'}(r)\right]
	\end{eqnarray}
	are the matrix elements related to the BdG wave functions and the temperature dependent functions $\chi$ and $\zeta$ are given by,
	\begin{eqnarray}
	\chi_{nn'}(\omega) &=& \Bigg(\frac{1}{\omega+i0^+-E_n-E_{n'}}
	-\frac{1}{\omega+i0^++E_n+E_{n'}}\Bigg)\lt(1-F_n(T)-F_{n'}(T)\rt), ~~\text{and}\no\\ \zeta_{nn'}(\omega) &=& \Bigg(\frac{1}{\omega+i0^++E_n-E_{n'}}
	-\frac{1}{\omega+i0^+-E_n+E_{n'}}\Bigg)\lt(F_n(T)-F_{n'}(T)\rt).
	\end{eqnarray}
\end{widetext}

It is useful to analyze the structure of $\chi$ and $\zeta$, since they occur in all the matrix elements of the inverse fluctuation propagators and provide insight about the microscopic processes that control the temperature dependence of the pair spectral function. Here $\zeta$ represents (upto matrix elements, which do not change its singularity structure) the probability amplitude of scattering a Bogoliubov quasiparticle from one state to the other. Note that $F_n(T)=0$ for all gapped states at $T=0$ and hence this term does not contribute to the collective modes around the ground state. A simple way to understand this is that the quasiparticles need to be present in the first place to be scattered, and at $T=0$, none of the gapped modes are excited in the system. As temperature is raised, this amplitude becomes finite. It is important to note that the singularities of this function occur when $\omega =E_n -E_{n'}$, and hence at very low energies. Thus at finite temperatures, $\zeta$ is complex at low energies, with an amplitude that increases with temperature. We will later see that these scattering processes play a very important role in determining the low energy pair spectral function at finite temperatures. We now consider the structure of $\chi$, which represents (upto matrix elements, which do not change the singularity structure of these functions) the amplitude for creating a pair of Bogoliubov quasiparticles. This is reflected in the singularities at $\omega= \pm (E_n+E_{n'})$. Hence,  for $\omega < 2 E_{gap}$, where the fermionic single-particle gap $E_{gap}$ corresponds to the lowest positive eigenvalue of the BdG Hamiltonian, $\chi$ is purely real, while it takes complex values for $\omega > 2 E_{gap}$. If we consider the numerator of $\chi$, it is evident that the numerator goes to $1$ at $T=0$. So the $T=0$ spectral function is completely dominated by this term. As the temperature is raised, the numerator decreases; however $\chi$ remains real at low energies below the two-particle continuum as long as the single-particle gap remains finite. 

The inverse fluctuation propagator for the phase fluctuation, $D^{-1}_{22}$ is given by,
\begin{eqnarray}
\nonumber D^{-1}_{22}(r,r',\omega) &=& \tilde{D}_{dia}(r,r')+ 
\omega^2\kappa(r,r',\omega) + \Lambda(r,r',\omega), \\
\end{eqnarray}
where $\tilde{D}_{dia}$ is the diamagnetic response of the system, $\kappa$ is the frequency dependent compressibility and $\Lambda(r,r',\omega)$ is related to the paramagnetic current-current correlator on the lattice. The exact formulas for $\tilde{D}_{dia}$, $\kappa$ and $\Lambda$ are given in Appendix~\ref{app:Fluct}.

\begin{widetext}
	Finally, the inverse fluctuation propagator corresponding to amplitude-phase mixing, $D^{-1}_{12}$ is given by
	\begin{eqnarray}
	{D^{-1}}_{12}(r,r',\omega) 
	&=& -\frac{i\omega}{4} \sum_{E_{n,n'}>0} f^{(1)}_{nn'}(r)f^{(2)}_{nn'}(r') \chi_{nn'}(\omega)+ f^{(1)}_{nn'}(r')f^{(2)}_{nn'}(r) \zeta_{nn'}(\omega).
	\end{eqnarray}
\end{widetext}
We invert the matrix $D^{-1}_{\alpha\beta}(r,r',\omega)$ to obtain the fluctuation propagators $D_{\alpha\beta}(r,r',\omega)$ and the corresponding spectral functions, ${\cal P}_{\alpha\beta}(r,r',\omega)=-\frac{1}{\pi} \text{Im} D_{\alpha\beta}(r,r',\omega)$. Here ${\cal P}_{11}(r,r',\omega)=
-\frac{1}{\pi} \text{Im}\langle
\eta(r,\omega+i0^+)\eta(r',-\omega+i0^+)\rangle$ corresponds to amplitude or Higgs fluctuations, ${\cal P}_{22}(r,r',\omega)=
-\frac{1}{\pi} \text{Im}\langle
\theta(r,\omega+i0^+)\theta(r',-\omega+i0^+)\rangle$  denotes the phase fluctuations while the amplitude-phase mixing is governed by ${\cal P}_{12}(r,r',\omega)=
-\frac{1}{\pi} \text{Im}\langle
\eta(r,\omega+i0^+)\theta(r',-\omega+i0^+)\rangle$.
However, the phase fluctuation propagators are not directly measureable in experiments, where probes couple to the electron density or current. As shown in Ref.~\onlinecite{HiggsAbhisek}, the experimentally measureable pair spectral function
\begin{eqnarray}
P(r,r',\omega) &=& \sum_{\alpha\beta}P_{\alpha\beta}(r,r',\omega),
\end{eqnarray}
where $P_{11} ={\cal P}_{11}$, $P_{12}(r,r',\omega) = \Delta_0(r){\cal P}_{12}(r,r',\omega)$, 
$P_{21}(r,r',\omega) = \Delta_0(r'){\cal P}_{21}(r,r',\omega)$, and 
$P_{22}(r,r'\omega) = \Delta_0(r)\Delta_0(r'){\cal
	P}_{22}(r,r',\omega)$. 

Note that in a translation invariant system, $P$ and ${\cal P}$ are related by simple scaling factors. However in a disordered system, where the pairing amplitude $\Delta_0(r)$ is varying in space, the spatial correlations of ${\cal P}$ and $P$ will be quite different and hence it is important to study the physically measureable correlations.

\begin{figure*}
	\includegraphics[width=0.95\textwidth]{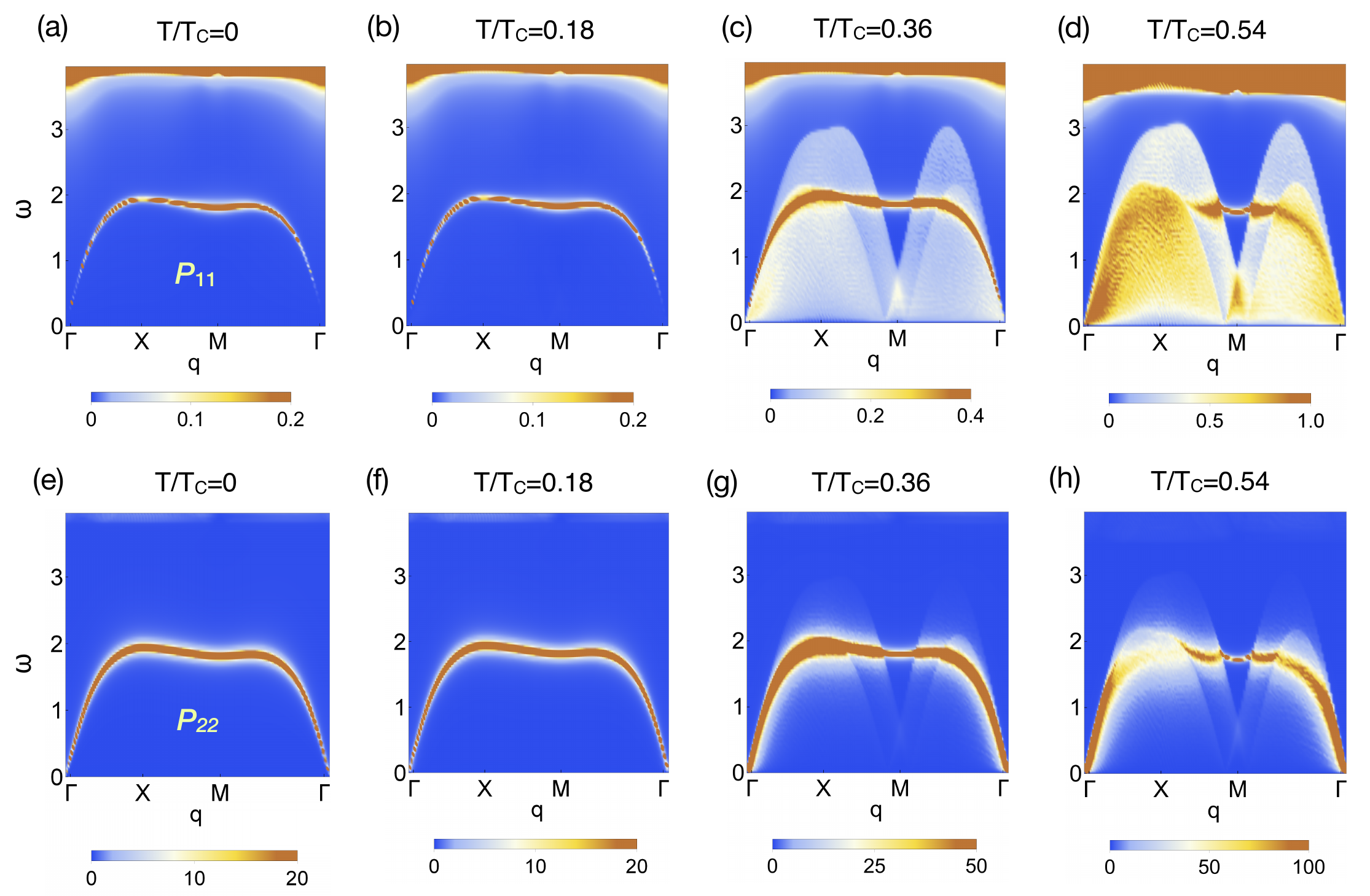}
	\caption{ Spectral functions in clean superconductor: (a)-(d) amplitude spectral function $P_{11}$ and (e)-(h) phase spectral function $P_{22}$ of the pair spectral function $P(q,\omega)$ in a clean superconductor ($V/t=0$) shown as a density plot in $q$ (along the principle axis of the 2D Brillouin zone of the square lattice) and $\omega$ for increasing temperatures: $T=0, T=0.18~T_c, T=0.36~T_c$ and $T=0.54~T_c$ (corresponding to $T=0, T=0.2~t, T=0.4~t$ and $T=0.6~t$ respectively). At small temperatures, both the spectral functions $P_{11}$ and $P_{22}$ consist of the collective modes below two-particle continuum. The temperature dependent background halo is clearly seen in $P_{11}$ for $T=0.36~T_c$ and $T=0.54~T_c$, while $P_{22}$ is mostly dominated by strong collective modes. 
	All data for the clean superconductors have been obtained on a $100\times100$ square lattice.}
	\label{PV0}
\end{figure*}
\subsection{Pair spectral function in a clean superconductor}
In this paper, we are primarily interested in studying the temperature dependence of the collective modes and the resultant two-particle spectral functions for a disordered superconductor. We start with the behavior of the temperature dependence of the two-particle spectral function $P(q,\omega)$ in the clean limit ($V/t=0$). This allows us to interpret the low energy spectral functions in terms of a temperature broadened collective mode and a background spectral weight arising from the scattering of thermally excited quasiparticles. This framework will then be used to investigate the pair spectral functions in the disordered case.

For a clean system, the problem simplifies considerably since the fluctuation propagators are diagonal in the momentum basis; e.g.
\begin{widetext}	
	\begin{equation}
	{D^{-1}}_{11}(q,\omega) = \frac{1}{U} + \frac{1}{2}\sum_{k}
	\lt[f^{(1)}_k(q)\rt]^2 \chi_k(q,\omega) + 
	\lt[{f^{(2)}_k(q)}\rt]^2 \zeta_k(q,\omega),
	\end{equation}
	where	
	\begin{eqnarray}
	f^{(1)}_k(q) &=&  u_ku_{k+q}-v_kv_{k+q},~~f^{(2)}_k(q) =  u_kv_{k'}+v_ku_{k'},
	\end{eqnarray}
	with 
	\begin{eqnarray}
	\chi_k(q,\omega) &=& \lt(\frac{1}{(\omega+i0^+-E_k-E_{k'})}
	-\frac{1}{(\omega+i0^++E_k+E_{k'})}\rt)(1-F_k(T)-F_{k'}(T))\\
	\zeta_k(q,\omega) &=& \lt(\frac{1}{(\omega+i0^++E_k-E_{k'})}
	-\frac{1}{(\omega+i0^+-E_k+E_{k'})}\rt)(F_k(T)-F_{k'}(T)).
	\label{eq:zeta}
	\end{eqnarray}
\end{widetext}
%
In the above formulae, we have used the standard BCS spectrum $E_k=\sqrt{\xi_k^2+\Delta_0^2}$~ with $\xi_k=-2t(\cos k_x + \cos k_y) - \mu$, $\Delta_0$ the uniform pairing amplitude and $u_k^2=1/2(1+\xi_k/E_k)=1-v_k^2$.

\begin{figure}
	\includegraphics[width=0.95\columnwidth]{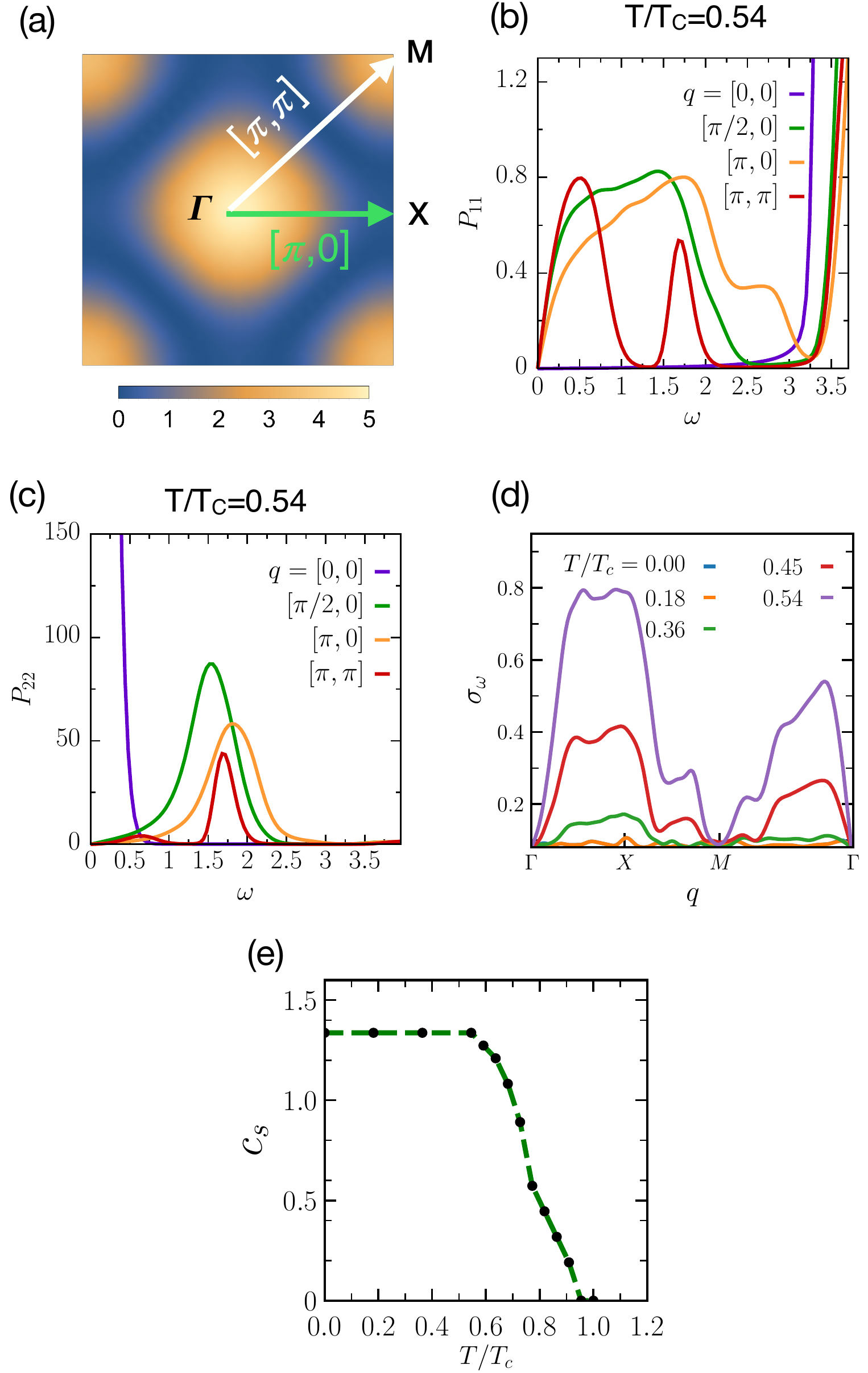}
	\caption{ (a): Color plot of the dispersion of a clean superconductor in the square lattice Brillouin zone. The wave-vector $[\pi,\pi]$, shown with white arrow, connects momenta where the dispersion is almost same. This leads to a low threshold for incoherent spectral weight due to quasiparticle scattering near $[\pi,\pi]$. On the other hand, the wave-vector $[\pi,0]$ connects momenta with large difference in dispersion. Hence the energy threshold for incoherent weight is large near $[\pi,0]$. (b) and (c): The EDC curves for (b) $P_{11}$ and (c) $P_{22}$ are shown for specific $q$ values at $T=0.54~T_c$. 
	(d): Width of the dispersive collective modes in $\omega$, ($\sigma_{\omega}$, obtained from the phase sector $P_{22}$) as a function of $q$ values  for different temperature values of $T$. We notice that that collective mode remains sharp at $\Gamma$ and $M$ points, while it broadens with temperature at intermediate momenta.
	(e): The sound velocity is plotted as a function of $T/T_c$. It remains almost constant up to $T/T_c\sim0.54$, then rapidly decreases to zero near $T_c$.
	}
	\label{Cleanfig}
\end{figure}

Fig.~\ref{PV0} (a)-(d) shows the amplitude spectral function $P_{11}(q,\omega)$, while Fig.~\ref{PV0} (e)-(h) shows the phase spectral function $P_{22}(q,\omega)$ in the clean system with increasing temperature. Here the attractive interaction $U=5~t$ and the density is set to 0.875. Let us first focus on the pair spectral functions at $T=0$ (Fig.~\ref{PV0} (a) and (e)). There is diffuse continuum spectral weight above $\omega > 2\Delta_0$, corresponding to propagation of two Bogoliubov quasiparticles. Note that at $T=0$, the $\zeta$ terms do not contribute, while $\chi$ is complex only for $\omega > 2\Delta_0$. For $\omega < 2\Delta_0$, there is a coherent dispersing peak at the collective mode frequencies determined by the vanishing of the determinant of the inverse fluctuation propagator. The mode disperses linearly at low momenta. The collective mode is a mixture of amplitude and phase fluctuations at finite momenta, but reduces to a pure phase Goldstone mode as $q \rightarrow 0$. As the temperature is raised to $T=0.18~T_c$ (Fig.~\ref{PV0} (b)), $T=0.36~T_c$ (Fig.~\ref{PV0} (c)) and $T=0.54~T_c$ (Fig.~\ref{PV0} (d)), a thermally broadened collective mode is clearly present riding on a distinct background halo. 

The background halo, which is due to the scattering of the quasiparticles already present in the system (the $\zeta$ terms), increases in intensity with increasing temperature. This incoherent spectral weight has some interesting characteristics. At each $q$, there is an upper bound of energy beyond which there is no incoherent spectral weight, till one reaches $\omega=2 \Delta_0$. This limiting energy, which is the maximum of $|E_{k}-E_{k+q}|$ for a fixed $q$, disperses linearly at small $q$ and shows a sharp dip around the $\Gamma$ and $M$ ($[\pi,\pi]$) points. In Fig.~\ref{Cleanfig} (a) we plot the dispersion of the quasiparticle energy $E_{k}$ as a function of $k$ in the Brillouin zone. We see that the wavevector $q=[\pi,\pi]$ only connects points in the Brillouin zone where the values of $E_k$ do not differ much, leading to a dip in the temperature dependent background halo around the $M$ point. Although only one such connection is shown in the figure, one can easily see that this is true in general. The same also holds for the $\Gamma$ point. On the contrary, the wave-vector  $q=[\pi,0]$ can connect points in the Brillouin zone where the values of $E_k$ can vary from a small to large value, and hence the lobe like structure extends up to a large value of $\omega$. As a result one can see that the collective mode both at $q=0$ and $q=[\pi,\pi]$ remain sharp, while there is considerable broadening at intermediate momenta. This is clearly seen in Fig.~\ref{PV0} (c) and (d) where the apparent width of the collective mode shrinks near the $M$ point when the collective mode lies above the band of incoherent spectral weight. This is also shown in Fig.~\ref{Cleanfig} (b), where we plot the line cuts of the Higgs spectral function along the energy axis (EDC or energy distribution curve) for fixed values of momenta at the largest temperature $T=0.54~T_c$. Near the zone center, the spectral weight lies above the two particle continuum. As we move along the $q_x$ axis, the modes at $[\pi/2,0]$ and $[\pi,0]$ do not show up as sharp peaks due to the large background incoherent weight. However, at $[\pi,\pi]$, one can clearly see two bumps in the spectral function, the lower one coming from the incoherent scattering of quasiparticles and the upper one corresponding to the coherent collective mode in the system.

We note that the background halo is more clearly seen in the Higgs spectral functions, while the phase spectral functions (Fig.~\ref{PV0} (e)-(h)) are primarily dominated by the large collective mode peak. The robust linear dispersion of the phase mode allows us to extract the speed of sound from the long wavelength linear dispersion. This speed of sound $c_s$ is plotted as a function of $T/T_c$ in Fig.~\ref{Cleanfig} (e). We see that at low temperatures $c_s$ remains almost constant, whereas near $T/T_c\sim0.54$ it starts decreasing and drops to zero at $T_c$. The background spectral weight in the phase sector is clearly seen only around $T=0.54~T_c$ (Fig.~\ref{PV0} (h)), where the characteristics are similar to that of the Higgs spectral weight. The EDC curves for the phase spectral function at fixed momenta are plotted in Fig.~\ref{Cleanfig} (c). Here it is clear that the coherent spectral weight in the collective mode is much larger than the incoherent spectral weight. Hence, near the collective mode frequency, one can expand the phase spectral function $P_{22}(q,\omega)\sim Z(q)/(\omega-\omega(q) +i \sigma_\omega(q))$. The large coherent spectral weight in the phase channel allows us to extract an energy width of the peak, $\sigma_\omega$ from the line cuts in Fig.~\ref{Cleanfig} (c). This width is plotted as a function of momenta for different temperatures in Fig.~\ref{Cleanfig} (d). It is clear that the collective mode remains sharp at $q=[0,0]$ and $[\pi,\pi]$, while the broadening at intermediate momenta increases with increasing temperature. 


\begin{figure*}
	\centering
	\includegraphics[width=0.95\textwidth]{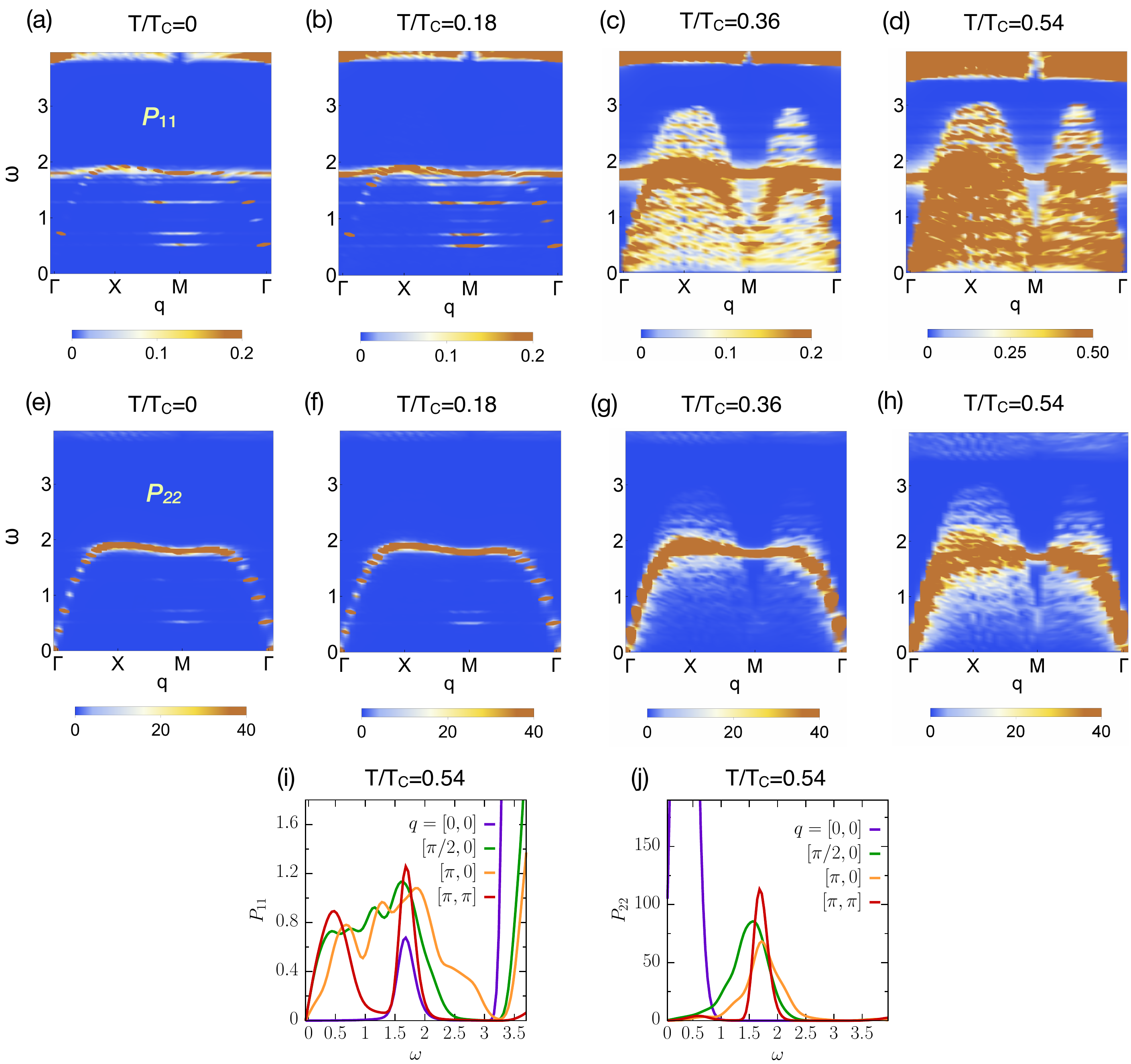}
	\caption{ Spectral functions in a weakly disordered superconductor: (a)-(d) amplitude spectral function $P_{11}$ and (e)-(h) phase spectral function $P_{22}$ of the pair spectral function $P(q,\omega)$ in the presence of weak disorder $V/t=0.1$ shown as a density plot in $q$ and $\omega$, for increasing temperatures: $T=0, T=0.18~T_c, T=0.36~T_c$ and $T=0.54~T_c$ (corresponding to $T=0, T=0.2~t, T=0.4~t$ and $T=0.6~t$ respectively). Note the appearance of the subgap Higgs peak in the amplitude sector $P_{11}$ at low $T$. While this non-dispersive Higgs mode remains unaffected at momenta $\Gamma=[0,0]$ and $M=[\pi,\pi]$ with increasing $T$, it gets overwhelmed by the temperature induced background halo at other momenta. The coherent collective modes in the phase sector remain mostly unaffected at low temperatures, but they are thermally broadened at large temperatures. 
		The EDC curves for (i) $P_{11}$ and (j) $P_{22}$ are shown, for specific $q$ values and for a temperature $T=0.54~T_c$. The EDC of $P_{11}$ clearly shows the Higgs mode at the $\Gamma$ point and two distinct modes (the lower broad peak is due to the temperature induced quasiparticle scattering, and the higher energy peak is due to disorder) at the $M$ point. The EDC curves for $P_{22}$ show thermally broadened dispersive collective mode peaks. All the disorder data have been obtained on a $24\times24$ lattice, and averaged over 15 disorder realizations.}
	\label{PV0p1}
\end{figure*}
\begin{figure*}
	\includegraphics[width=0.95\textwidth]{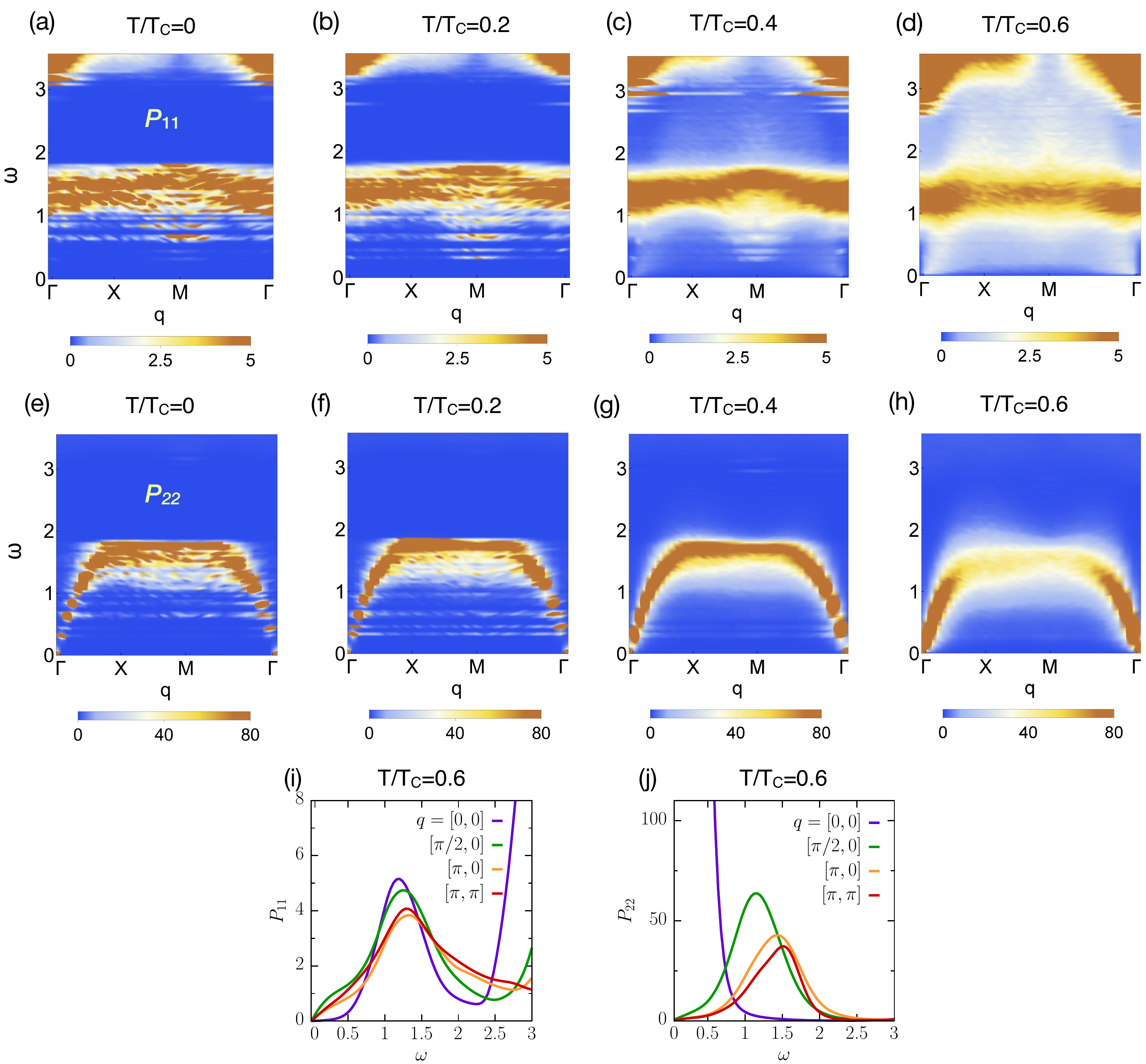}
	\caption{ Spectral functions in a moderately disordered superconductor: (a)-(d) amplitude spectral function $P_{11}$ and (e)-(h) phase spectral function $P_{22}$ of the pair spectral function $P(q,\omega)$ in the presence of moderate disorder $V/t=1.0$ shown as a density plot in $q$ and $\omega$, for increasing temperatures: $T=0, T=0.2~T_c, T=0.4~T_c$ and $T=0.6~T_c$ (with $T_c=1.0$). The subgap Higgs mode gets broadened and its low energy threshold comes down in energy. With increase in temperature, contrary to the weak disorder, the lobe structure of the background halo is not seen in amplitude sector, and the Higgs mode remains prominent at all temperatures. The EDC curves for (i) $P_{11}$ and (j) $P_{22}$ are shown for specific $q$ values and for $T=0.6~T_c$. The EDC curves for $P_{11}$ also show that the non-dispersive Higgs mode dominates at all temperatures. On the other hand, the phase sector is dominated by the dispersing collective modes, which only get thermally broadened at large temperatures.}
	\label{PV1p0}
\end{figure*}
\begin{figure*}
	\centering
	\includegraphics[width=0.85\textwidth]{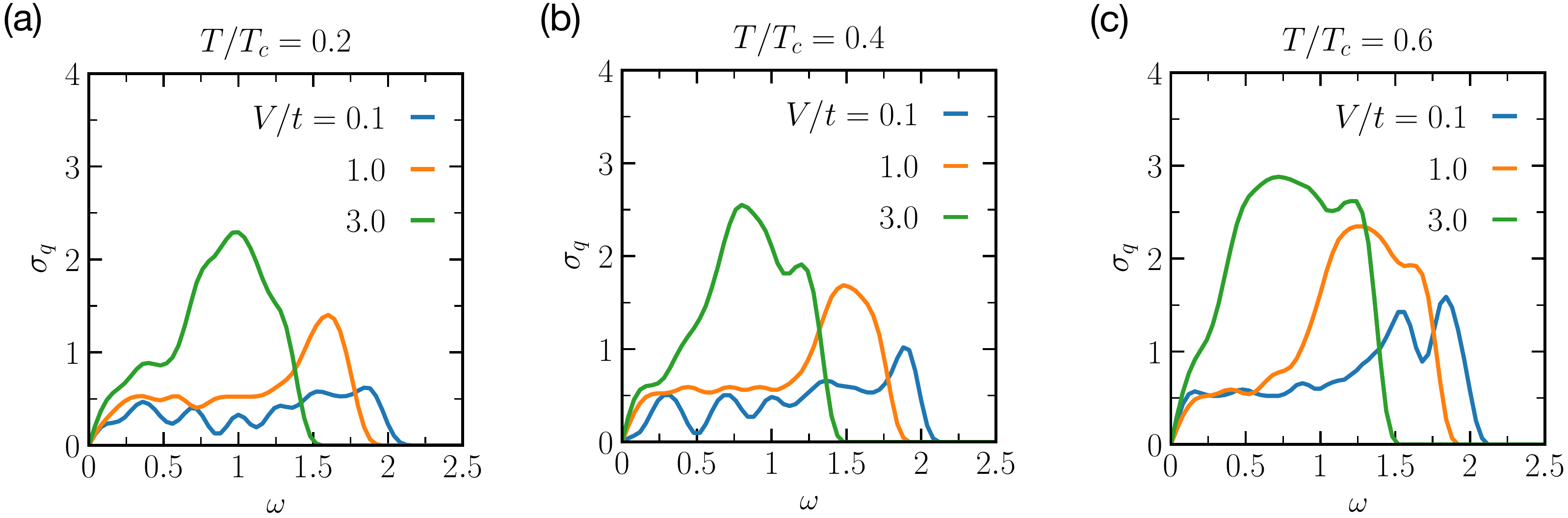}
	\caption{ (a)-(c): Width of the collective modes in $q$, ($\sigma_q$, obtained from the phase spectral function $P_{22}$) as a function of $\omega$ below two particle continuum, for (a) $T=0.2~T_c, T=0.4~T_c$ and $T=0.6~T_c$ respectively. We note that for a fixed $T$, $\sigma_q$ increases with increase in disorder. At  small disorder the width does not change much with energy, while the larger disorder shows a broad peak. Since with increasing disorder the collective mode structure comes down to lower energy (Fig.~\ref{PV0p1} and \ref{PV1p0}), the threshold $\omega$ at which $\sigma_q$ vanishes also decreases with disorder.}
	\label{qwidth}
\end{figure*}

\subsection{Pair spectral function in disordered superconductor}
We now consider the key issue which we want to study in this paper: how do the collective modes evolve with temperature in a disordered superconductor. In a disordered system, momentum is not a good quantum number for a single disorder realization. For each such realization, we first construct the pair spectral function $P_{\alpha\beta}$ as a matrix in the real space co-ordinates $r$ and $r'$. We then work with the center of mass co-ordinate $R=(r+r')/2$ and relative co-ordinate $d=(r-r')$, and average the spectral function $P(d,R,\omega)$ over several disorder realizations. The disorder averaging restores translation invariance, i.e. the averaged quantity is a function only of $d$ and $\omega$. Averaging over $R$, we get $P(d,\omega)=1/N_s \langle \sum_R P(d,R,\omega)\rangle$ ($N_s$ being the number of lattice sites and $\langle\rangle$ corresponds to disorder average). We then Fourier transform the spectral function in $d$ to express it as a function of $q$ and $\omega$, i.e. $P(q,\omega)$. The variation of this disorder averaged spectral function with momentum $q$ and energy $\omega$ will be our key tool to study the behaviour of finite temperature collective modes.

We first consider the spectral function of the disordered system at $T=0$ (worked out in Ref.~\onlinecite{HiggsAbhisek}), so that we have a reference to understand the finite temperature variations. The amplitude and phase spectral functions $P_{11}$ and $P_{22}$ are plotted as a function of $q$ and $\omega$ for a weakly disordered system with $V=0.1~t$ in Fig.~\ref{PV0p1} (a) and (e) respectively. While the phase spectral function (Fig.~\ref{PV0p1} (e)) is almost unchanged from the clean case, with a linearly dispersing collective mode dominating at low energies, the amplitude spectral function shows dramatic change (Fig.~\ref{PV0p1} (a)). In contrast to the clean case, where at $q=0$, the Higgs mode sits at the threshold of the two-particle continuum at an energy of $2\Delta_0$, a non-dispersive mode appears at an energy below two-particle continuum ($2E_{gap}$) in this case. At $q=0$ this subgap mode is identified as the disorder-induced Higgs mode in a superconductor~\cite{HiggsAbhisek}. The $T=0$ spectral functions for a moderately disordered system with $V=~t$ is shown in Fig.~\ref{PV1p0} (a) (amplitude) and (e) (phase) respectively. The non-dispersive mode in the amplitude spectral function gains more spectral weight and is considerably broadened, while the phase spectral function is relatively unchanged with disorder.

Next we study the effect of disorder on the amplitude and phase spectral functions at finite temperatures. Fig.~\ref{PV0p1} (b)-(d) shows the amplitude spectral function $P_{11}(q,\omega)$, and Fig.~\ref{PV0p1} (f)-(h) shows the phase spectral function in presence of a weak disorder $V/t=0.1$ with increasing temperature. 
The most visible change in the amplitude spectral functions is the appearance of the low energy continuum weight or the halo in the background of the collective mode. As explained in the section on clean superconductors, this weight represents the scattering of the Bogoliubov quasiparticles and increases with temperature. However, for each $q$ there is an upper bound of energy up to which the background weight exists. This background cut-off disperses linearly at small $q$ and shows a pronounced dip around the $[\pi,\pi]$ point. Therefore the non-dispersing Higgs mode near  $q=0$ and $q=[\pi,\pi]$ remains unaffected at finite temperatures (Fig.~\ref{PV0p1} (b), (c) and (d)). We also note that the small Higgs component in the linearly dispersing collective mode is overwhelmed by the background incoherent weight as temperature increases (Fig.~\ref{PV0p1} (c) and (d), corresponding to $T/T_c=0.36$ and $T/T_c=0.54$), so the only coherent weight in the amplitude spectral function at these finite temperatures is related to the disorder-induced Higgs mode. To see this feature clearly, we plot some energy distribution curves (EDCs) of the amplitude spectral function at $T=0.54~T_c$ in Fig.~\ref{PV0p1} (i). These are line-cuts of the data in Fig.~\ref{PV0p1} (d) at fixed values of $q$. The non-dispersive mode at $q=0$ is clearly seen as a peak. At $q=[\pi,\pi]$, there are two peaks, with the lower broad peak corresponding to the incoherent quasiparticle scattering, and the sharper higher energy peak (at energies similar to the $q=0$ peak) corresponding to the non-dispersive Higgs mode.


On the other hand, the phase spectral function $P_{22}$ remains mostly unaffected in the presence of weak disorder even at finite temperatures, as seen in Fig.~\ref{PV0p1} (f)-(h). While the collective mode is thermally broadened, it still dominates the low energy phase spectral function function. It is interesting to note that the mode near $[\pi,\pi]$ remains sharp at finite temperatures as the incoherent spectral weight lies below the energy of this mode. As temperature is increased (Fig.~\ref{PV0p1} (g) and (h)), the background halo with the two lobe structure becomes more prominent even in the phase spectral function. The variation of the phase spectral function with energy at fixed momenta at $T=0.6~T_c$ is plotted in Fig.~\ref{PV0p1} (j). The curves show thermally broadened dispersive peaks of the collective modes.

\begin{figure*}
	\centering
	\includegraphics[width=0.85\textwidth]{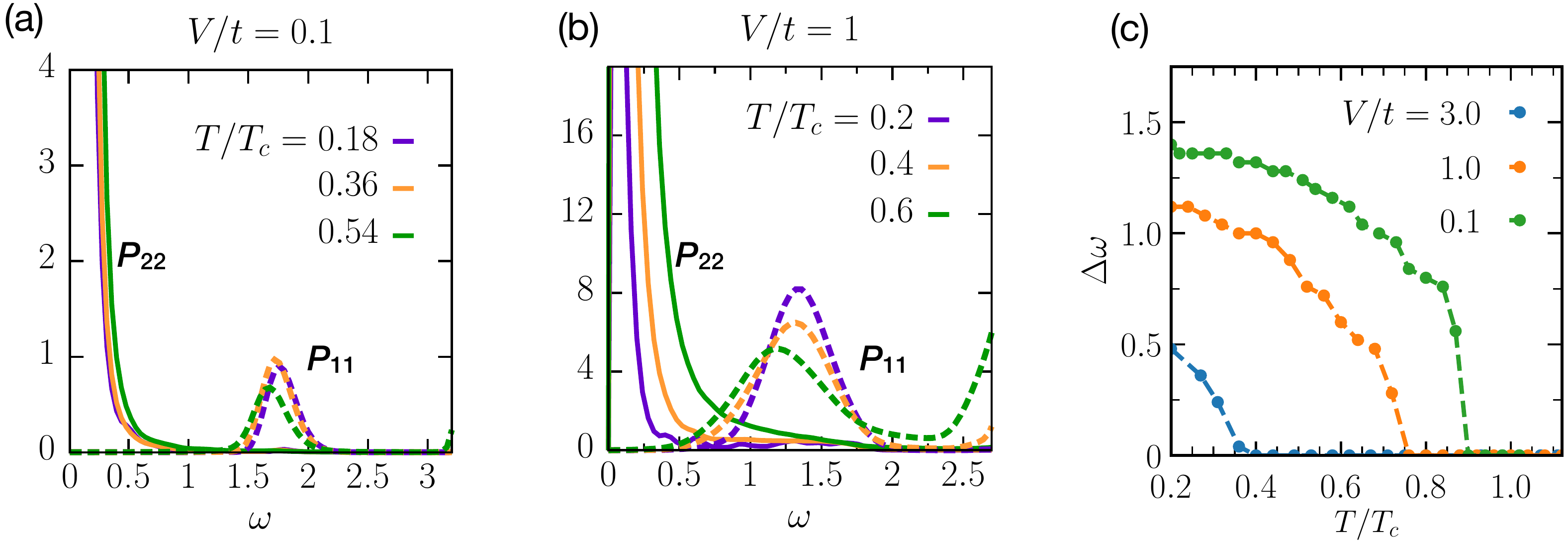}
	\caption{ (a)-(b): Amplitude ($P_{11}$, shown as dotted lines) and phase contribution ($P_{22}$, shown as solid lines) to the pair spectral function $P$ at $q=[0,0]$ shown for two different disorder values, $V=0.1~t$ and $V=t$, as a function of $\omega$, and for increasing temperatures. (c): The separation between the Higgs and the phase peak, $\Delta\omega$ plotted as a function of $\omega$. We notice that at small temperatures, they are separated for moderately large values of disorder, while it falls rapidly to zero at a critical temperature $T_c^s$ with increase in disorder.}
	\label{Phiggs}
\end{figure*}

We now increase the disorder to a moderate value of $V=~t$ and study the spectral functions with increasing temperature. With increase in disorder, the non-dispersive Higgs mode gets broadened and its lower end comes down towards the zero energy. This mode also gains much more spectral weight. This is seen in Fig.~\ref{PV1p0} (a) where we plot the spectral function at $T=0$. As temperature is increased (Fig.~\ref{PV1p0} (b)-(d)), we once again see a diffuse background halo, but at this moderate disorder, the sharp lobe structure of the halo, which was present in the clean and the weakly disordered system, is absent. Since the sharp boundaries resulted from simultaneous momentum and energy conservation in quasiparticle scattering, and momentum conservation is strongly broken in each disorder realization at these moderate disorders, the background halo is more diffuse in this case. However, as in the weakly disordered case, the non-dispersive Higgs mode remains prominent at all finite temperatures. This is clearly seen in Fig.~\ref{PV1p0} (i), where we plot the variation of the amplitude spectral function with energy at fixed momenta. The curves at all the momenta show a broad peak at roughly the same energy corresponding to the non-dispersive Higgs mode in the system. We note that at this moderate disorder, the edge of the continuum perceptibly comes down with increasing temperature, showing the softening of the gap in the system.

In contrast, the phase spectral function is dominated by the linearly dispersing collective mode, which is robust to both the presence of disorder and temperature (Fig.~\ref{PV1p0} (e)-(h)). At the largest temperature of $T=0.6~T_c$, the collective mode near the $[\pi,0]$ or $[\pi,\pi]$ point is broadened, but a distinct peak can still be observed, as seen in the EDCs plotted in (Fig.~\ref{PV1p0} (j)).

We have already seen that the phase spectral function consists of a dominant dispersing collective mode. However in a disordered system, momentum is not a good quantum number, and one would expect the collective modes to be broadened in momentum space due to elastic scattering from the impurities. To estimate the effect of this scattering, we consider the half width of the spectral function peak in the phase channel at different fixed values of $\omega$ from the momentum distribution curves (MDCs).  We only take into account the collective mode line between $\Gamma$ and $X$ points where the mode is dispersing, and limit our study to energies well below two particle continuum. In Fig.~\ref{qwidth} (a)-(c) we plot this width $\sigma_q$ as a function of $\omega$ for three different temperatures, $T=0.2~T_c$, $T=0.4~T_c$ and $T=0.6~T_c$ respectively. Each plot contains the width for three different disorder values, a weak disorder of $V=0.1~t$, a moderate disorder of $V=t$ and a strong disorder of $V=3~t$ respectively. As expected, we observe that for any fixed  $T$, $\sigma_q$ increases with increasing disorder. While the low disorder width does not change much with energy of the collective modes, the width at moderate and high disorders show a broad peak as a function of collective mode frequency. We also find that $\sigma_q$ vanishes at a threshold $\omega$ which decreases with disorder. This happens because the whole collective mode structure itself comes down when we increase disorder (see Fig.~\ref{PV0p1} and \ref{PV1p0}).




We have already seen that the non-dispersive mode in the amplitude channel produces finite subgap spectral weight at $q=0$, while the linearly dispersing collective mode has large weight at zero energy in the phase channel (the Goldstone mode). The phase peak and the amplitude peak are spectrally separated at $T=0$, and hence this mode should be spectroscopically observable. We now consider whether a finite temperature will erase this spectral separation and render this mode invisible. We have plotted the amplitude and phase contribution to the spectral function for a weak disorder of $V/t=0.1$ (Fig.~\ref{Phiggs} (a)) and a moderate disorder of $V=t$ (Fig.~\ref{Phiggs} (b)). In both these cases, we find that as temperature is increased, the peak positions remain unchanged while the broadening increases, but the separate phase and amplitude features are observable upto a reasonably high temperature. Thus this feature is also robust to turning on temperature in the system. We note that inclusion of density fluctuations can alter the spectral separation of these features~\cite{densityfluct}.


To systematically track the separation between the subgap Higgs peak and the low energy phase peak, we define a parameter $\Delta\omega$ which indicates to the separation between them in energy. In Fig.~\ref{Phiggs} (c), we plot $\Delta\omega$ as a function of temperature for disorder $V/t=0.1, 1$ and 3. Here we extend our analysis up to large temperature values, keeping in mind that the BdG theory does not work well close to $T_c$. We notice that at small temperature, the Higgs and the phase modes are separated for moderately large value of disorder. However, with increase in temperature, the separation decreases monotonically and vanishes at a critical temperature $T^s_c$. $T^s_c$ decreases with increase in disorder, which suggests that the sharp feature of the Higgs mode is more robust in presence of temperature at small disorder and the robustness goes away with increase in disorder. The momentum and energy resolved MEELS spectroscopy~\cite{Abbamonte} should observe this Higgs mode separately from the phase pileup in an energy resolved way.

In conclusion, in this work we have extended our previous studies on two-particle spectral function for disordered $s$-wave superconductors~\cite{HiggsAbhisek} to finite temperatures. Using a functional integral formalism and gaussian expansion around the inhomogeneous saddle point, we have studied the two-particle spectral function at small and moderately high temperatures, both in clean and disordered superconductors. We derive the analytical formulas for inverse fluctuation propagators at finite temperature, continued to real frequency. We present the full $q-\omega$ dependence of the amplitude and phase sectors of the spectral function, and therefore study the evolution of the \textit{Higgs} and the Goldstone mode with temperature and disorder. We show that at finite temperatures, additional low energy incoherent spectral weight appears in the form of lobes. In presence of disorder, these temperature dependent background halo competes with the collective modes in the amplitude sector. However, we find that if the disorder is not too strong, the non-dispersive Higgs mode which appears as a subgap feature at $q=[0,0]$ remains unaffected in presence of moderately high temperatures. Therefore, the Higgs mode can be seen in an energy resolved way separately from the low energy phase pile-up, even at experimentally accessible temperatures.
\begin{acknowledgements}
	A.S and R.S. acknowledge the computational facilities of the Department of Theoretical Physics, TIFR Mumbai. N.T. acknowledges support from DOE grant DE-FG02-07ER46423.
\end{acknowledgements}

\appendix
\section{Superfluid stiffness at finite temperature}
\label{app:SFstiff}
We use Bogoliubov transformation in a disordered superconductor, which diagonalizes the effective mean-field Hamiltonian for the negative $U$ Hubbard model, with energy $E_n$ and the corresponding eigenfunction $[u_n(r),v_n(r)]$~\cite{Nandini1}.
Here $n$ runs over the positive eigenvalues i.e. $E_n>0$.
The current operator is defined as
\begin{equation}
j^x_r = it\sum_{\sigma} \lt( c^\dagger_{r+\hat x\sigma}c_{r\sigma} - c^\dagger_{r\sigma}c_{r+\hat x\sigma}\rt),
\end{equation}
and the local kinetic energy associated with the $x$-directed hopping is given by
\begin{equation}
K^x_r = -t\sum_{\sigma} \lt( c^\dagger_{r+\hat x\sigma}c_{r\sigma} + c^\dagger_{r\sigma}c_{r+\hat x\sigma}\rt).
\end{equation}
Now the superfluid stiffness by the Kubo formula is given by,
\begin{eqnarray}
{D_s\over \pi} = \langle - K^x\rangle -\Lambda_{xx}(q_x=0,q_y\rightarrow 0,i\omega_p=0),
\end{eqnarray}
where $i\omega_p$ is the Bosonic Matsubara frequency. The first term represents the diamagnetic response to an external magnetic field which is given by,
\begin{equation}
\langle -K^x\rangle\! =\! {4t\over N} \sum_{n}\!\! u_n(r+\hat x)u_n(r)F_n(T)+v_n(r+\hat x)v_n(r)(1-F_n(T))
\label{kx}
\end{equation}
The second term is the paramagnetic response given by the dynamical transverse current-current correlation function,
\begin{equation}
\Lambda_{xx}({\bf q},i\omega_p) = {1\over N}\int_0^{1/T}\!\! d\tau\ e^{i\omega_p \tau} \langle j^x({\bf q},\tau),j^x(-{\bf q},0)\rangle
\end{equation}
which is calculated to be,
\begin{widetext}
	\begin{eqnarray}
	\Lambda_{xx}({\bf q},i\omega_p) &=& {2t^2\over N}\sum_{nm} {A_{nm}({\bf q})(A_{nm}({\bf q})+B_{nm}({-\bf q})) \over i\omega_p+(E_n-E_m)} \lt(F_n(T)-F_m(T)\rt).
	\end{eqnarray}
\end{widetext}
In the above equation, $n$ and $m$ run over all eigenvalues (both $E_n<0$ and $E_n>0$), and $A_{nm}$ and $B_{nm}$ are given by,
\begin{equation}
A_{nm} ({\bf q}) = \sum_{i} e^{-i{\bf q}.r} (u_n(r+\hat x)u_m(r)-u_n(r)u_m(r+\hat x)) \no
\end{equation}
\begin{equation}
D_{nm} ({\bf q}) = \sum_{i} e^{-i{\bf q}.r} (v_n(r+\hat x)v_m(r)-v_n(r)v_m(r+\hat x)). \no
\end{equation}
\section{Inverse fluctuation propagator for  phase fluctuations}
\label{app:Fluct}
The inverse fluctuation propagator for the phase fluctuation is given by,
\begin{eqnarray}
\nonumber D^{-1}_{22}(r,r',\omega) &=& \tilde{D}_{dia}(r,r')+ 
\omega^2\kappa(r,r',\omega) + \Lambda(r,r',\omega).
\end{eqnarray} 
The diamagnetic response $\tilde{D}_{dia}$ is related to the local kinetic energy
\begin{equation}
\nonumber {\cal K}(r,r') = -t\sum_{E_n>0} v_n(r)v_n(r')[1-F_n(T)] + u_n(r)u_n(r')F_n(T)
\end{equation} 
through the relation
\begin{equation}
\tilde{D}_{dia}(r,r') = -\frac{1}{2} \delta_{rr'}\sum_{\langle rr_1\rangle }{\cal K}(r,r_1)+\frac{1}{2}\delta_{\langle rr'\rangle}{\cal K}(r,r')
\end{equation}

\begin{widetext}
	Here, the frequency dependent compressibility $\kappa$ is given by density density correlator, 
	\begin{eqnarray}
	\kappa(r,r',\omega) 
	&=& \frac{1}{8}\sum_{E_{n,n'} >0}
	f^{(2)}_{nn'}(r)f^{(2)}_{nn'}(r')\chi_{nn'}(\omega) +
	f^{(1)}_{nn'}(r)f^{(1)}_{nn'}(r')\zeta_{nn'}(\omega),
	\end{eqnarray} 
	while $\Lambda(r,r',\omega)$ is related to the paramagnetic current-current correlator on the lattice 
	\begin{eqnarray}
	\Lambda(r,r',\omega) &=& \sum_{\langle r r_1\rangle \langle r' r_2\rangle}
	J(r,r_1,r',r_2,\omega) - J(r,r_1,r_2,r',\omega)
	- J(r_1,r,r',r_2,\omega) + J(r_1,r,r_2,r',\omega)~~~~ \\ \no \\
	\text{where} \quad\quad J(r,r_1,r',r_2,\omega) &=&
	- \frac{t^2}{8} \sum_{E_{nn'}>0} f^{(3)}_{nn'}(r,r_1)f^{(3)}_{nn'}(r_2,r')\chi_{nn'}(\omega) + f^{(4)}_{nn'}(r,r_1)f^{(4)}_{nn'}(r_2,r')\zeta_{nn'}(\omega).
	\end{eqnarray}
	The new matrix elements $f^{(3)}$ and $f^{(4)}$ are given by
	\begin{eqnarray}
	f^{(3)}_{nn'}(r,r') &=& \left[u_n(r)v_{n'}(r')-v_n(r)u_{n'}(r')\right],~~\text{and}~~f^{(4)}_{nn'}(r,r')= \left[u_n(r)u_{n'}(r')+v_n(r)v_{n'}(r')\right].
	\end{eqnarray}
\end{widetext}

\bibliographystyle{unsrt}
\bibliography{refs.bib}

\end{document}